\documentclass[aps,twocolumn,prl,preprintnumbers]{revtex4}

\begin{document}

\preprint{MCTP-06-24}
\preprint{NSF-KITP-06-123}
\preprint{hep-th/0612056}

\title{Chain Inflation via Rapid Tunneling in the Landscape}
%\date{December 7, 2006}

\author{Katherine Freese}
\email{ktfreese@umich.edu}
\affiliation{Michigan Center for Theoretical Physics,
University of Michigan, Ann Arbor, MI 48109--1040}

\author{James T.~Liu}
\email{jimliu@umich.edu}
\affiliation{Michigan Center for Theoretical Physics,
University of Michigan, Ann Arbor, MI 48109--1040}
\affiliation{Kavli Institute for Theoretical Physics,
University of California, Santa Barbara, CA 93106}

\author{Douglas Spolyar}
\email{dspolyar@physics.ucsc.edu}
\affiliation{Physics Department, University of California,
Santa Cruz, CA 95060}

\begin{abstract}

Chain inflation takes place in the string theory landscape as the
universe tunnels rapidly through a series of ever lower energy vacua
such as may be characterized by quantized changes in four form fluxes.
The string landscape may be well suited to an early period
of rapid tunneling, as required by chain inflation, followed
by a later period of slow tunneling, such as may be required
to explain today's dark energy and small cosmological constant.
Each tunneling event (which can alternatively be thought of as a
nucleation of branes) provides a fraction of an $e$-folding of
inflation, so that hundreds of tunneling events provide the requisite
amount of inflation.  A specific example from M-theory compactification
on manifolds with non-trivial three-cycles is presented.

\end{abstract}

\maketitle

%%%%%%%%%%%%%%%%%%%%%%%%%%%%%%%%%%%%%%%%%%%%%%%%%%%%%%%%%%%
In 1981, Guth \cite{guth} proposed an inflationary phase of the
early universe to solve the horizon, flatness, and monopole problems
of the standard cosmology.  During inflation, the Friedmann
equation
\begin{equation}
H^2 = 8\pi G_4 \rho /3 + k/a^2
\end{equation}
is dominated on the right hand side by a (nearly constant) false
vacuum energy term $\rho \simeq \rho_{vac} \sim$ {\it constant}.  The
scale factor of the Universe expands superluminally, $a \sim t^p$ with
$p>1$.  Here $H$ = $\dot a /a$ is the Hubble parameter. With
sufficient inflation, roughly 60 $e$-folds, the cosmological
shortcomings are resolved.  Inflationary models fall into two classes:
tunneling models such as Guth's original model and rolling models
where a scalar field slowly rolls down an extremely flat potential.

In Chain Inflation \cite{fs}, the universe tunnels rapidly through a series of
vacua separated from one another by energy barriers.  One can imagine
a multidimensional potential containing many ``bowls'' or minima of
varying energies. The universe starts out in a high-energy minimum,
and then sequentially tunnels down to bowls of ever lower energy until
it reaches the bottom.  During the time spent in any one of these
bowls, the universe inflates by a fraction of an $e$-fold.  After many
hundreds or thousands of tunneling events, the universe has inflated
by the requisite 60 (or so) $e$-folds to resolve the cosmological
problems outlined above, as well as satisfying $\delta\rho/\rho$ \cite{fink}.  At each stage, the phase transition is rapid
enough that percolation is complete, though thermalization and
reheating does not take place until the final few stages of tunneling
near the bottom of the potential.  In principle the energy scales of
the potential can vary widely, from GeV scale up to almost Planck
scale. No fine-tuning of the shape of the potential is required (the
potential need not be flat).

A simple way to visualize Chain Inflation is 
to imagine a tilted cosine potential (sometimes known as a ``washboard
potential'') as the effective path through the multidimensional
landscape.  Then the universe tunnels from a high energy minimum of
the cosine down to the next minimum and so on.  In fact, we showed in
\cite{fls} that the QCD axion, which has been proposed as a byproduct
of a solution to the strong CP problem in the theory of strong
interactions, can have a tilted cosine potential and hence is a
candidate for the inflaton in chain inflation; in this case inflation
takes place at the QCD scale and can in principle be experimentally tested
in the foreseeable future.

In this paper we focus instead on chain inflation in the string
landscape. Chain inflation is quite generic in the string
landscape, as the universe tunnels through a series of metastable vacua,
each with different fluxes%
\footnote{KF thanks N. Arkani-Hamed for drawing attention to this fact.}.
There has been a great deal of
interest in the fact that the vacuum of string theory is not unique;
in fact there appear to be at least $10^{200}$ different vacua.  Even
with compactification on a specific Calabi-Yau manifold,
there is still a large variety of background fluxes that could be
turned on, providing a huge number of different possible vacua.  These
vacua with different fluxes are disconnected in the mutidimensional
landscape, with barriers in between them. Thus the only way to proceed from
one to the other is via tunneling.  Chain inflation is the result of
rapid tunneling through a series of these vacua.

{\it Chain inflation.} In chain inflation,
the universe begins in some metastable minimum of the potential with a
positive value of vacuum energy.  From this minimum
there would presumably be a variety of possible directions in field space
in which the system could tunnel to a lower minimum.
However, many of the neighboring minima
would be inaccessible because the tunneling rate into them is too slow;
instead, the system would tunnel more quickly in a different direction.
The system thus chooses the ``path of least resistance'', {\it i.e.} it
moves through a series of minima to which the tunneling is the most rapid.

From a given initial minimum, tunneling will proceed to the neighboring
minimum with the largest drop in energy, lowest barrier height and
smallest difference in field value.  In our first paper, (hereafter
Paper 1) \cite{fs}, we quantified this statement by modeling any single
tunneling event using a scalar field with an asymmetric 
double well potential
\begin{equation}
V(\phi) = {1 \over 4} \lambda (\phi^2 - a^2)^2
+ {\epsilon \over 2a} (\phi - a).
\end{equation}
Then the tunneling rate is determined by three quantities: the
barrier height $\lambda a^4/4$, 
energy difference between vacua $\epsilon$, and 
width of the potential $2a$.
We noted that, for the case where all three scales are comparable,
the tunneling rate is on the borderline of being extremely rapid
or extremely slow relative to a Hubble time. The reason for this
extreme sensitivity to parameters is the exponential dependence
of the tunneling rate on the bounce action, $\Gamma \sim e^{-B}$
where (in the thin-wall approximation) \cite{callan,coleman,deLuccia}
\begin{equation}
B=\frac{27\pi^2}2\frac{\tau^4}{|\epsilon|^3}.
\label{eq:baction}
\end{equation}
Here $\epsilon$ is the drop in energy and $\tau$ is the tension
of the brane interpolating between the two vacua.  In particular,
for the above $\lambda\phi^4$ potential, the brane tension is given
by $\tau=\frac{2\sqrt2}3\sqrt\lambda a^3$.  Hence, for
natural parameter choices,
the tunneling can be either very fast or very slow.  The first choice of
the field will be to move in the fastest direction of large $\epsilon$,
small barrier height, and small width $a$.

{\it The string landscape.} A typical picture of the string landscape
involves a complicated distribution of numerous metastable minima
with very small energy differences and separated by large ({\it i.e.}~string
scale) barrier heights.  This view of the landscape has been driven
in part by attempts to solve the cosmological constant problem, in which
case one is naturally driven to consider energy differences on the
order of the presently observed vacuum energy,
$\epsilon \sim (10^{-3}\,{\rm eV})^4$.  From this point of view, it
is generally expected that tunneling would be extremely slow.  However,
we note that, although metastable minima with very small energy differences
are naturally present in the landscape, they would generically be well
separated from each other.  Instead, adjacent minima
are much more likely to have large energy differences for the simple
reason that the natural scale for a single configuration change
({\it e.g.} the nucleation of a D-brane instanton) is given by the string
scale, and {\it not} by something like $10^{-3}\,{\rm eV}$.  Since
both energy differences and barriers between adjacent minima in the
landscape are expected to be characterized by the same string scale,
this indicates that the tunneling rate may be quite large.  Furthermore,
Tye recently
noted another mechanism for fast tunneling \cite{tye}: even in situations
where the tunneling between neighboring minima is typically slow,
resonant effects may greatly enhance the tunneling rates.

Portions of the landscape can be quite well suited to an early period
of rapid tunneling, as required by chain inflation, followed
by a later period of slow tunneling, such as may be required
to explain today's dark energy and small cosmological constant.
As we noted in Paper 1, given a choice of tunneling to a number
of lower energy vacua of different energies, the field will take
the largest steps first.  The field will tunnel in the direction
of the larger vacuum energy difference, as the tunneling in that direction
is fastest. For example, if the field has a choice of tunneling
into a minimum with energy difference $\epsilon \sim (0.1\, m_{pl})^4$
or a minimum with energy difference $\epsilon \sim (10^{-10}\, m_{pl})^4$,
it will choose the former.  The path taken by the field quite
naturally lends itself to an early period of chain inflation with
rapid tunneling.  Subsequently, the field slows down, as the only
remaining minima it can tunnel to have smaller and smaller energy
differences from its current position.

We also note that chain inflation is {\it not} a model of eternal
inflation.  During the rapid tunneling phase, the rate of tunneling
back up the potential is very small.  
Chain inflation is a dynamical  model in which
the field chooses the downward
path of most rapid tunneling.  Hence, the universe is
driven universally to one or at most a few final states, rather
than a huge number of them.   
Given any minimum the field is sitting in,
there may be more than one rapid path, but not a huge number of them.
This arbitrariness of path must be considered when evaluating density
perturbations in the model.  One may need to invoke the anthropic
principle to explain our current universe, but only to a small degree.

{\it Tunneling and the cosmological constant.}
While chain inflation was initially formulated in terms of scalar
fields \cite{fs}, it may equally well occur in any landscape which
allows rapid tunneling through a chain of metastable minima.  Since
typical string models populate the landscape by turning on various
fluxes, it is natural in this case to have flux-driven chain inflation
(and in particular four form flux when viewed in $3+1$ dimensions).
Here we will
see that the energy difference between minima indeed decreases as one
goes to lower energy vacua, so that we can have the tunneling rate
change from initially being rapid to subsequently becoming smaller.
It is possible that this change in tunneling rate leads from
a period of chain inflation to a universe stuck in its current
small value of the cosmological constant.

To compute the tunneling, we follow the lead of Brown and Teitelboim
\cite{bt1,bt2} (hereafter BT).  The BT mechanism may be motivated
by the observation that, in 1+1 dimensional electrodynamics, an
electron/positron pair can be produced in the presence of a sufficiently
strong background electric field.  The particles are physically pulled
apart by the original field,
and the region between them has a reduced value of electric
field. The world lines of the electron and the positron can be thought
of as membranes (domain walls), with the electric fluxes in the
regions interior to the membranes reduced by a quantized amount.  One
can think of this process as the nucleation of membranes in the
presence of background electric fields. A series of nucleation events
takes place, each time reducing the value of the electric flux by a
quantized amount. 

As in BT, we are interested in the four-dimensional generalization of
this process, in which background fields cause bubble nucleation of
lower energy regions with lower field values.  While the BT mechanism
relies on four-form fluxes in $3+1$ dimensions, such fluxes have numerous
potential origins in the string theory landscape.  For example, in the
case of IIB flux vacua, both NSNS and RR three-form internal flux are
generically present.  Taking the Hodge dual results in seven-form flux,
which reduces on three-cycles of the internal Calabi-Yau manifold to
yield four-form fluxes in $3+1$ dimensions.  This flux may then be
reduced through the nucleation of both wrapped NS5-branes and wrapped
D5-branes.  Alternatively, from an M-theory (or IIA) point of view,
the M-theory four-form can take values in either spacetime or internal
space, with corresponding flux reduced through the nucleation of either
M2-branes or wrapped M5-branes, respectively.  In all such cases, the
nucleated branes manifest themselves as membranes in the $3+1$ dimensional
spacetime.

The nucleation of domain walls changes the value of the vacuum
energy by giving rise to quantized drops in the value of
the background fluxes; four-form
field strengths play the role of a dynamical vacuum energy
in four dimensional spacetime,
\begin{equation}
\label{eq:cc}
\Lambda = \Lambda_{\rm bare} +\frac12F_4^2,
\end{equation}
where $\Lambda_{\rm bare}$ is the bare cosmological constant, and
$F_{\mu\nu\rho\sigma}=F_4\epsilon_{\mu\nu\rho\sigma}$, so $F_4$ is the
magnitude of a four-form field which can change its value through the
nucleation of membranes.  The change of the field strength across the
membrane is
\begin{equation}
\label{eq:change}
\Delta F_4 = \pm q,
\end{equation}
where the charge $q$ is constant so that the flux values are quantized,
at least in string theory.  Hence, Eq.~(\ref{eq:cc}) becomes
\begin{equation}
\label{dont}
\Lambda = \Lambda_{\rm bare} + {1 \over 2} n^2 q^2,
\end{equation}
where the integer $n$ labels the integer flux.

At each stage of domain wall production, the integer flux changes from
$n \rightarrow n-1$, and the vacuum energy changes by an amount
\begin{equation}
\label{eq:epsilon}
\epsilon = -(n-1/2) q^2.
\end{equation}
This is the energy difference between vacua in chain inflation;
one can think of this process as tunneling from one vacuum to 
another whose energy is lower by the amount given in Eq.~(\ref{eq:epsilon}).

Bousso and Polchinski (hereafter BP) extended the BT approach to
include multiple four-form fluxes \cite{bp}.  They noted that the
quantized energy changes in Eq.~(\ref{eq:epsilon}) with a single flux
in BT are too large to end up with a small vacuum energy today of
order the dark energy, $\Lambda \sim 10^{-120} m_{pl}^4$.  On the
other hand, compactification manifolds could easily have a large
number of cycles, and turning on fluxes on all these cycles leads
to many effective four-form fluxes in $3+1$ dimensions.  By using
a large number $J$ of fluxes with a $J$-dimensional grid of
vacua, they noted that the spectrum of allowed values of $\Lambda$ can
then be sufficiently dense to explain the observed value of the
cosmological constant in some portion of the universe.  We will follow
their multi-flux proposal, but dispense with subscripts that differentiate
between one flux and another (e.g. we will use the notation $n$ to
generically mean the integer flux for any one of the four forms).
Furthermore, while BP had the goal of neutralizing the cosmological constant
in this fashion, we are more directly interested in chain inflation in
the landscape.  Hence we do not necessarily concern ourselves with the
details of the endpoint of the BP scenario.

{\it Requirements for chain inflation.}
In the zero temperature limit, the nucleation rate $\Gamma$ per unit
spacetime volume for producing bubbles of true vacuum in the sea of
false vacuum through quantum tunneling can be expressed as
\begin{equation}
\label{eq:tunrate}
\Gamma(t) = A e^{-S_E} = A e^{-(B-S_{BG})}.
\end{equation}
Here the Euclidean action $S_E$ is the difference between the
instanton action $B$ given (in the thin-wall limit) by Eq.~(\ref{eq:baction})
and the background Euclidean action of de~Sitter
space $S_{BG} = - {8 \pi^2 /H^2}$.  $A$ is a determinantal factor
which can be estimated as $A \sim 1$ \cite{gar,vil}.  As seen in
Eq.~(\ref{eq:baction}), the instanton action depends on the difference
in energy $\epsilon$ between the inside and outside the bubble and
on the domain wall tension $\tau$. Chain inflation consists of a number of
rapid tunneling steps, in which case the bubbles are small compared to
the horizon size.  In this case, the gravitational
correction $S_{BG}$ can be ignored, so that $S_E \sim B$.  Later
we will comment on a possible treatment beyond the
thin-wall limit.

%\footnote{We
%  note that we do not need to include gravitational effects
%  \cite{deLuccia} as they would only be relevant for bubbles
%  comparable to the horizon size, whereas the bubbles produced here
%  are much smaller.}. 

Guth and Weinberg \cite{guthwein} have shown that the
probability of a point remaining in a false de~Sitter vacuum is
approximately
\begin{equation}
\label{eq:probds}
p(t) \sim \exp\left({-{4 \pi \over 3}\beta H t}\right),
\end{equation}
where the dimensionless quantity $\beta$ is defined by
$\beta \equiv \Gamma/H^4$.
The number of $e$-foldings per tunneling event is
\begin{equation}
\label{eq:ni}
\chi = \int H dt \sim H \tau ={3 \over 4\pi\beta}. 
\end{equation}
The authors of \cite{guthwein,tww} calculated that a critical value
of
\begin{equation}
\label{eq:betacrit}
\beta \geq \beta_{\rm crit} = 9/4\pi
\end{equation}
is required in order to achieve percolation and thermalization.
In terms of number of $e$-foldings, this is
\begin{equation}
\label{eq:chicrit}
\chi \leq \chi_{\rm crit} = 1/3.
\end{equation}
As long as this is satisfied, the phase transition at each stage takes
place quickly enough so that `graceful exit' is achieved. Bubbles
of true vacuum nucleate throughout the universe at once, and
are able to percolate. 
Sufficient inflation requires the total number of $e$-foldings to
satisfy 
\begin{equation}
\label{eq:totreq}
N_{\rm tot}>60.
\end{equation}
Old inflation \cite{guth} failed in that a single tunneling event can satisfy
either Eq.~(\ref{eq:chicrit}) or Eq.~(\ref{eq:totreq}) but not both.
Chain inflation satisfies both equations by having a chain of tunneling
events, each of which provides only a fraction of an $e$-folding but
adding up to a total of at least 60-folds.

{\it Chain inflation in string theory.}
Given the above requirements, we now examine the feasibility of realizing
chain inflation in the string landscape picture.  In general, fluxes in
string theory are closely related to branes, in that $p$-branes are
natural sources for $(p+2)$-form fluxes.  The model we have in mind
is essentially that of BP \cite{bp}, where we start with a $D$ dimensional
theory (where $D$ may be 10 or 11) compactified down to $3+1$ dimensional
spacetime.  The objects that we are interested in are BPS $p$-branes
which satisfy a $D$ dimensional relation between charge and tension
\begin{equation}
\mu_p^2=2\kappa_D^2\tau_p^2.
\label{eq:Dbps}
\end{equation}
For D$p$-branes we have $\tau_p=(\sqrt\pi/\kappa_{10})
(2\pi\sqrt{\alpha'})^{3-p}$,
while for M2 and M5-branes, we have instead $\tau_2=2\pi M_{11}^3$
and $\tau_5=2\pi M_{11}^6$, respectively, where
$2\pi M_{11}^9=1/(2\kappa_{11}^2)$.

In order to obtain four-form fluxes in $3+1$ dimensions, we consider
$p$-branes wrapped on $(p-2)$-cycles of the compactification manifold.
In this case, the effective brane tension in
four dimensions is given simply by
\begin{equation}
\tau=\tau_p V_{p-2},
\label{eq:4tension}
\end{equation}
where $V_{p-2}$ is the volume of the corresponding $(p-2)$-cycle which
the brane wraps.  In addition, the brane charge measured in four
dimensions can be shown to be
\begin{equation}
q=\frac{\mu_p V_{p-2}}{\sqrt{V_{D-4}}},
\label{eq:4charge}
\end{equation}
where $V_{D-4}$ is the volume of the compactification manifold.
Using Eqs.~(\ref{eq:4tension}) and (\ref{eq:4charge}), we see that
the $D$-dimensional BPS relation, Eq.~(\ref{eq:Dbps}), reduces to the
universal four-dimensional relation
\begin{equation}
\tau=\frac1{\sqrt2}m_{\rm pl} q,
\label{eq:4bps}
\end{equation}
where the four-dimensional Planck mass ($m_{\rm pl}=2.43 \times 10^{18}$~GeV)
is related to the $D$-dimensional one by
\begin{equation}
\label{eq:M11}
m_{\rm pl}^2\equiv\frac1{8\pi G_4}= \frac{V_{D-4}}{\kappa_D^2}.
\end{equation}
Although we view this model of $(p+2)$-form fluxes reduced to four form
fluxes on various cycles as a toy model of the string landscape, we believe
that it captures many of the generic features common to any realistic flux
compactification setup.  In particular, we expect that the effective
tensions and charges are given, at least within the correct order of
magnitude, by Eq.~(\ref{eq:4bps}).

Starting for a given metastable minimum specified by a particular set of
fluxes, we now estimate the tunneling rate of Eq.~(\ref{eq:tunrate}), or
equivalently, the membrane nucleation rate.  Although a generic state
in the landscape involves many independent fluxes (specified by a set
of integers $\{n_i\}$), because they
are independent (at least to leading order), it is sufficient to consider
the nucleation of a single membrane, which changes the flux of one of the
four-forms from $n$ to $n-1$.  In this case, the change in the vacuum energy
is given by Eq.~(\ref{eq:epsilon}). Taking $n\gg1$, we have
\begin{equation}
\label{eq:epsilon2}
|\epsilon| \approx n q^2 .
\end{equation}
For $n\gg1$, and for $\Lambda$ sufficiently large%
\footnote{Since chain inflation occurs prior to neutralization of the
cosmological constant, there is no contradiction between this assertion
and the framework of BP.},
gravity can be neglected so the bounce action is essentially given by
Eq.~(\ref{eq:baction}).  Inserting $\tau$ and $\epsilon$ corresponding
to the nucleated membrane into this expression yields \cite{bt1,bp}
\begin{equation}
\label{eq:tunaction}
B_n = { 27 \pi^2 m_{\rm pl}^6\over 16 n^3 \tau^2},
\end{equation}
where we have used the BPS relation, Eq.~(\ref{eq:4bps}), to relate the
brane charge $q$ with its tension $\tau$.
We note that the semiclassical approximation in which the instanton
calculations have been computed is valid for $B_n\gg1$.  Thus the above
expression cannot be trusted in the fast tunneling regime that we are
interested in (where $B_n\sim1$).  Nevertheless, the qualitative behavior
of $B_n$ as a function of the brane tension ought to remain valid, and
that is mainly what we are interested in.

It ought to be apparent from Eq.~(\ref{eq:tunaction}) that tunneling between
adjacent minima is governed by both the amount of flux $n$ along a given
cycle as well as the tension $\tau/m_{\rm pl}^3$ of the corresponding brane
(in Planck units) that must be nucleated to reduce this flux.  Note,
in particular, that tunneling is enhanced for larger brane tensions.
The reason behind this somewhat counterintuitive result is that the
change in energy $\epsilon$ increases as $\epsilon\sim\tau^2$, and this
is more than sufficient to compensate for the increased brane tension.
As far as this toy model is concerned, these adjacent minima are well
separated in energy, and hence fast tunneling is very much generic in
this landscape.  Metastable minima with tiny differences in energy (as
envisioned in BP) are well separated, and are only connected through
multiple flux jumps along many simultaneous cycles.  Tunneling between
such states are highly suppressed.  However, this is irrelevant, as
the system naturally tunnels through the most favorable path at any
given stage in the chain.  In the simple case where all brane tensions
are identical, the cycle with the largest $n$ ({\it i.e.}~supporting the
most flux) is the one which preferentially undergoes tunneling.  In
this case, at the beginning of chain inflation, the system is in a
state with fluxes specified by an initial set of independent quanta
$\{n^{\rm initial}_i\}$, but can naturally end up on the BP circle with
$\{n^{\rm final}_i\approx n_0\}$.

Of course, just because the system undergoes a series of tunneling events
until it dynamically neutralizes $\Lambda_{\rm bare}$ does not in itself
ensure that chain inflation actually occurs in the landscape.  To see
whether chain inflation is possible, we first compute the quantity
$\beta=\Gamma/H^4$ using Eq.~(\ref{eq:tunrate}) for the tunneling rate
as well as the Hubble relation $H^2 = \Lambda/3m_{\rm pl}^2$.
We find
\begin{equation}
\label{eq:betam}
\beta_n \equiv \frac\Gamma{H^4} = 9\left(\frac{\Lambda_n}{m_{pl}^4}\right)^{-2}
e^{-B_n},
\end{equation}
where $\Lambda_n$ is given by Eq.~(\ref{dont}) generalized to multiple fluxes
\begin{equation}
\Lambda_n=\Lambda_{\rm bare}+m_{\rm pl}^{-2}\sum_in_i^2\tau_i^2,
\label{eq:genlam}
\end{equation}
and we have introduced the subscript $n$ to indicate that the quantities
are dependent on the amount of flux that is threading the cycle that undergoes
this stage of tunneling.

At any step in the chain, $\beta_n$ must satisfy the inequality
\begin{equation}
\label{eq:betamm}
\beta_n\geq\frac9{4\pi}\approx0.716,
\end{equation}
given by Eq.~(\ref{eq:betacrit}) in order for percolation to succeed at each
tunneling transition.  From Eq.~(\ref{eq:betam}), this puts a requirement
on the bounce action
\begin{equation}
\label{eq:bouncreq}
B_n\alt2.53-2\log(\Lambda_n/m_{\rm pl}^4).
\end{equation}
Note that $\Lambda_n$ is the vacuum energy of the metastable minimum specified
by $n$ (actually the complete set $\{n_i\}$).  Depending on the parameters of
chain inflation, this can be
anywhere from near (but below) the Planck scale down to few tens of MeV's
(in order to generate a sufficient reheating temperature) \cite{fs}.
Furthemore, $\Lambda_n$ changes throughout the chain of tunneling
events, starting from some initial value $\Lambda_{\rm initial}$, and
dropping down to $\Lambda_{\rm final}\approx0$ at the end of inflation. 

Our final requirement is given by Eq.~(\ref{eq:totreq}), which ensures
that sufficient inflation occurs.  In particular,
\begin{equation}
\label{eq:suff}
N_{\rm tot} = \sum{3 \over 4 \pi \beta_n} \geq 60,
\end{equation}
where the sum is over the total number of tunneling events during
inflation.  As chain inflation proceeds, the value of $\beta_n$
given by Eq.~(\ref{eq:betam}) changes because of two competing effects.
The first is that, as $\Lambda_n$ is reduced, $\beta_n$ naturally
tends to increase.  However, the second is that, as the overall flux is
driven down, the values of $n$ decrease.  This leads to a larger
bounce action $B_n$, which in turn leads to an exponential suppression
of $\beta_n$.  Because $\beta_n$ is variable, Eq.~(\ref{eq:suff}) does
not lead itself to a simple expression for the total number of tunneling
events.  However, a necessary condition may be given as
\begin{equation}
\label{eq:deltan}
\hbox{\# of tunnelings} \ge 80\pi \beta_{\rm min}\ge180,
\end{equation}
where the second inequality is simply the percolation requirement of
Eq.~(\ref{eq:betamm}).

The constraints, Eqs.~(\ref{eq:bouncreq}) and (\ref{eq:deltan}), on
the bounce action and the number of tunnelings must be simultaneously
satisfied in chain inflation, using the effective four-dimensional brane
tension given by Eq.~(\ref{eq:4tension}).  For example, taking the M-theory
example of BP, we may consider the nucleation of both M2-branes and
M5-branes wrapped on three-cycles of the seven-dimensional compactification
manifold.  Inserting the appropriate brane tensions into
Eq.~(\ref{eq:tunaction}), we obtain
\begin{equation}
B_n=\frac{27}{64n^3}\left(\frac{m_{\rm pl}}{M_{11}}\right)^6
\times\cases{1&M2,\cr
(V_3M_{11}^3)^{-2}&M5.}
\end{equation}
In order to arrive at a large number of four-form fluxes, we consider
the case of wrapped M5-branes.  In this case, one example of a
satisfactory set of parameters is to take $M_{11} = 10^{-3}\, m_{\rm pl}$
along with $V_3M_{11}^3=10^3$.  Starting at $\Lambda_{\rm initial}
=(10^{-1}\,m_{\rm pl})^4$, we see that $\beta_n>1$ for $B_n<21$, which
is obtained for $n>2735$.  Assuming inflation ends at $\Lambda_{\rm final}
=(10^{-20}\,m_{\rm pl})^4$, a similar computation indicates that
$n>1045$ at the end of inflation.  Note that, for these parameters,
the effective wrapped brane tension is $\tau=2\pi\times 10^{-6}\,m_{\rm pl}^3$.
Eq.~(\ref{eq:genlam}) then indicates that each
flux contributes on the order of $4n^2\times10^{-11}\,m_{\rm pl}^4$ to
$\Lambda$.  For $n\sim10^3$, the energy difference per step is then on
the order of $\epsilon\sim10^{-7}\,m_{\rm pl}^4$.  As a result, about
1000 tunneling events will be needed to neutralize $\Lambda_{\rm initial}$.

The above requirements on $n$ (along with
that of a sufficient number of tunnelings) may be satisfied by
an appropriate choice of $\Lambda_{\rm bare}$ in the BP framework.
Note that such large values of flux may destabilize
the background manifold due to backreaction, or could prove difficult
to realize because of tadpole constraints.  However, this issue is
unresolved (see, {\it e.g.}, \cite{dewolfe,kumar}).
In this case, it might be better to consider the alternative
of large effective brane tension $\tau$ rather than large $n$ as responsible
for rapid tunneling.  In this M5-brane example, this corresponds to taking
a compactification manifold with large three-cycles.

{\it End of inflation.}
We have used four forms to illustrate a possible implementation
in M-theory of the basic ideas of chain inflation.
The nice feature exists that, as time goes on, the tunneling
rate automatically slows down as the integer units of flux $n$
decreases.  However, it is hard
to see how inflation ends. While the general
idea of rapid tunneling followed by slow tunneling is attractive,
the trouble is that there can be an intermediate regime
of medium tunneling.  This intermediate regime would mimic
the reheating problems of old inflation, in that the universe
could end up with too many empty big bubbles that have not
percolated.  However, this could potentially be avoided through
resonant tunneling \cite{tye} or devaluation \cite{deval}.
Another possible way to avoid this problem is to have a very
large Planck-scale negative value of $\Lambda_{\rm bare}$ that
gets cancelled off by the positive vacuum contributions we
have been discussing in this paper. Then the landscape would
involve a steep set of metastable minima (analogous to a linearly
tilted cosine potential) that persists into negative
values of the vacuum energy.  The naive thin-wall computation
of tunneling would give us rapid tunneling into AdS, which
would obviously be disastrous. However, it has been argued
by Banks and others \cite{banks} that tunneling into AdS is forbidden,
in which case chain inflation would simply stop at the last
positive value of vacuum energy before going negative.   

{\it Beyond thin-wall.} Our treatment of membrane nucleation as
an effective tunneling process for chain inflation has up to this
point assumed the thin-wall limit, where the energy difference between
barriers is much smaller than the barrier height.  The computation of
the instanton action in Eq.~(\ref{eq:baction}) by
\cite{callan,coleman,deLuccia} made this assumption.  Banks, however, noted
that one can go beyond the thin-wall computation \cite{banks}.
As mentioned previously, the tunneling process is the
higher dimensional analog of pair production in 1+1 dimensional
electrodynamics, the massive Schwinger model.  Using techniques
of bosonization, this model can be shown \cite{cosine} to be
equivalent to a scalar field theory with Lagrangian
\begin{equation}
\label{eq:cosine}
L = {1 \over 2} [(\Delta \phi)^2 - {e^2 \over \pi} \phi^2
+ M^2 {\rm cos}(\phi)] .
\end{equation}
In this formulation, pair production is described as vacuum decay
of the false vacua at minima of the cosine, which correspond
to backgrounds with fixed quantized values of electric flux.

Banks notes that in higher dimensions, though the bosonization 
does not generalize, one can still introduce a scalar field through
$p$-forms:
\begin{equation}
F_{\mu_1 ... \mu_p} = \epsilon_{\mu_1 ... \mu_p} \phi .
\end{equation}
He argues that a derivative expansion for the effective action for
$\phi$, with inclusion of effects of membranes, will again lead to the
form of Eq.~(\ref{eq:cosine}), though with the cosine replaced by a
periodic function.  In \cite{fls}, we studied chain inflation with a
tilted cosine potential similar to the one in Eq.~(\ref{eq:cosine}).
A treatment of tunneling beyond the thin-wall limit with a tilted
cosine potential is underway \cite{fls2}.

{\it Discussion.}
We have shown that chain inflation can take place in the string
theory landscape as the universe tunnels through a series of 
ever lower energy vacua. This is not a variant of eternal inflation,
as tunneling back up the potential is suppressed \cite{vil}.
We illustrated an example
in which these vacua are characterized by quantized changes in four-form
fluxes.  Although we have focused on effective $3+1$ dimensional
dynamics, these ideas could of course be generalized to a series
of tunneling events via $p$-forms in $p$-dimensions.  

We also
note that chain inflation can take place at low scales ({\it e.g.}~TeV)
and so can be compatible with low scale gravity $M_{11} \sim$ TeV 
as is the case for large extra dimensions.  Most single field rolling models
fail at such low scales, as they require potentials with Planck scale
width and GUT scale height.

A recent paper by de Alwis \cite{dealwis} claims that, using the
string effective supergravity action together with branes, there is no
tunneling.  However, this result was not derived in the full string theory.
Here we have assumed that the creation of M2-branes or M5-branes,
which is effectively a tunneling process, makes physical sense in an
estimate of the more precise theory.

\section*{Acknowledgements}

We would like to thank B. Acharya, T. Banks, K. Bobkov, R. Bousso,
O. DeWolfe, M. Dine, G. Kane, D. Marolf, L. McAllister, J. Polchinski and
S. Watson for many useful discussions.
KF thanks N. Arkani-Hamed for pointing out that chain inflation with
four forms is generic in the string landscape.  This work was supported
in part by the US Department of Energy under grant DE-FG02-95ER40899,
the National Science Foundation under grant PHY99-07949, and the
Michigan Center for Theoretical Physics under grant MCTP-06-27.
KF would like to thank the Miller Institute at the
University of California, Berkeley, for support and hospitality,
and JTL wishes to acknowledge the hospitality of the KITP.

\end{document}